Astro2020 Decadal Survey *State of the Profession* white paper

# A better consensus: Changes to the Decadal process itself

**David W. Hogg**
*New York University*
*Flatiron Institute*

**David Schiminovich**
*Columbia University*

Executive summary: The importance of the Decadal Survey in astrophysics is great; it deserves attention and revision. We make recommendations to increase the Survey's transparency and political legitimacy. The Astro2020 charge asks the Survey to "generate consensus recommendations". It is healthy to re-evaluate how to achieve consensus as the community and context evolve. Our recommendations are the following: (R1) **Appoint the Decadal panel chairs and panel members through a transparent process, or even a democratic process.** (R2) **Don't make panel members sign any kinds of non-disclosure agreements, or strictly limit these.** (R3) **Educate the community about the Decadal's decision-making and consensus-building.** (R4) **Provide written documentation about how white papers will be read and used.** (R5) **Give the community an opportunity to comment on and vote to approve the final reports.** (R6) **Ask the AAAC to help the agencies make these changes.**

The Decadal Survey process in US astrophysics has been extremely valuable to our community and especially the federal agencies that fund our research. Both anecdotally and comparatively to other disciplines, it is clear that the fact that astronomy speaks with a united voice through the Decadal Survey has had a huge influence on the resources available to us, the missions and projects funded for us, and the structures and organizations in which we get funding, work, and publish. The Decadal process is an important part of what we do as a community and it is important to our scientific successes, which have been legion in the last few decades.

Because this process is so important to the practice of astrophysics, it is also important that the process of reaching consensus—the consensus requested in the charge to the Decadal committee—itself be politically legitimate. Here we make specific recommendations that could improve the Decadal process. We make these recommendations in a completely constructive spirit. We realize that some of these recommendations are challenging to implement in the current structure. That is, this white paper is not just directed to the Decadal Survey panel, it is also directed to the National Academy of Sciences itself and to the funding agencies. But the recommendations are also incremental, in the sense that they don't involve any radical changes.

In recent years, astronomers have been struggling with myriad political and social issues, from pervasive sexual and racial harassment to matters of access and inclusivity. We have a long way to go to solve these problems, and to build the community that we want. The good news is that astronomers are openly facing the hard questions of our society and our profession, with a genuine commitment to bring changes. This commitment must be applied to the Decadal process itself. Making incremental changes to the Decadal (as we propose here) won't solve our hardest problems! But the Decadal Survey is too important to leave to the ways of old.

(R1) **Appoint the Decadal panel chairs and panel members through a transparent process, or even a democratic process.** An ideal Decadal panel will contain thought leaders and consensus builders. And panel members should be willing (at the end) to defend and advertise Decadal results to Congress, the agencies, and the community. Right now panel chairs (effectively) must be members of the NAS. The tacit assumption here is that membership in the NAS is desired. And yet, excellent astrophysicists nominated for NAS membership have been rejected. Once panel chairs are selected, they in turn (we infer) select panel members. These processes are opaque, accepting only nomination input; nominations are confidential and cannot be compared to final panel makeup.

Panel chair selection and panel member selection processes could involve some voting or election. Many European scientific advisory panels are populated through ballots, so it can work in practice. And there is research that shows that positions filled by competitive processes are better filled than those filled through expert nomination or appointment. Certainly there is no reason at present to believe that a ballot would produce a less capable or visionary committee than the current closed-door appointment process. If full elections are impossible, there could be an experiment of electing a subset of the chairs or committees or at-large members next round.

Even if the current Decadal panel-member selection process cannot be changed by NAS, some transparency about the metrics that are used to judge that the panels have been properly populated would be valuable. For example, what fraction of the panels should be early-career? What fractions from different kinds of institutions? And what fractions from different demographic categories? We aren't taking a position on what these targets should be; we are taking the position that these targets should be clearly stated and open for discussion and debate if the panels are to be considered representative for our community.

(R2) **Don't make panel members sign any kinds of non-disclosure agreements, or strictly limit these.** We have heard conflicting things about non-disclosure related to the NAS and the Decadal Survey. However, one of us (Hogg) was asked to sign a non-disclosure before joining the survey in 2009 (he refused, and was not permitted to serve). That non-disclosure agreement (attached as an appendix so there is no ambiguity) prohibited all discussion of closed-door discussions with no expiration date. A request (by Hogg in 2009) for an expiration of the NDA was rejected.

Non-disclosure is wrong; it prevents analysis or criticism of the process after the fact, and makes it difficult to recognize and fix problems. It is essential that the community be permitted to discuss what happened in each Decadal Survey and how the panels came to their recommendations. And NDAs are also a bad idea because they are easily violated by former panel members when they describe any matter that goes beyond what is in the publicly released reports.

Some may argue that confidentiality is required for a good process that is broad-reaching and for reaching consensus. A time-limited NDA might also serve a role in ensuring panel independence. But the benefit of a permanently closed-door process is not apparent. For example, the Astronomy and Astrophysics Advisory Committee (AAAC) that advises NASA, NSF, and DOE is a completely open committee (indeed anyone can come in by phone to any meeting, and no offline decision-making is permitted). Even with this structure in place, the AAAC has advised the agencies through some very difficult decisions.

(R3) **Educate the community about the Decadal's decision-making and consensus-building.** In our narratives, past Decadal Surveys are linked to the largest, most successful missions of the time. For example, the 1972 Decadal is given credit for the Hubble Space Telescope. And yet in that report the Large Space Telescope was not the top choice; it was Recommendation #9. What lesson should be drawn?

Neither the NAS nor the past Decadal panels have answered many basic questions about the process. Why have many good ideas have ended up lower ranked? How important is community popularity to a high ranking? How important is transformative promise? And how important are strategic considerations, with respect to (say) Congress, funding agencies, or space contractors? Are the Surveys officially risk-averse in primary recommendations? How should our scientists and engineers discuss and act on the final report? There is also confusion in the community about how much the Decadal Survey is about ranking large funding priorities (billion-dollar projects) and how much it is about guiding smaller funding programs (like individual grants programs) and how much it is about less financially relevant decisions (like publishing models or education). Although these S, M, L, XL questions can be answered in part by looking at reports from previous decades, the community has changed, the context has changed, and the panels have changed. It is not clear how similar Astro2020 will be to Astro2010, and how similar Astro2030 will be to either.

The Decadal provides an excellent opportunity to engage and educate the community on how our science is justified, funded, and communicated. And such engagement and education is necessary if we want all researchers from all backgrounds to be able to make critical decisions about how they spend their time in support of the process. We should not be afraid to openly acknowledge, understand, and question our community's approach to scientific decision-making and consensus-building, even if at first it may seem to take us further from our ultimate goals.

We recognize that this recommendation (R3) adds burden to the panels and the process: Our recommendations would increase cost and time. For example, the first acts of the panels would have to be educational and explanatory work, which would have to take place *before* the calls for community input.

(R4) **Provide written documentation about how white papers will be read and used.** An enormous effort is put into the writing of white papers. Many hundreds will be written, each by a team, each of which has spent many person-weeks. That corresponds to an expenditure of many millions of dollars equivalent of research time. And it may unfairly burden early-career scientists. <u>If it was clear what criteria are being used to decide what kinds of things in white papers end up in reports, we could potentially save our community substantial time and effort</u>. It would also lead to the submission of better and more useful white-paper input. Most of what we know about the usefulness and efficacy of white papers comes from explanations by former panel members (but see the discussion of recommendation R2). These explanations have been piecemeal, and unofficial. There ought to be a clear official position and rubric, one that gives non-trivial guidance. This is especially important to those early in their careers, who wonder about the strategic value of work they do to support white papers, relative to the value of advancing their research programs. Our recommendation is that the panels produce documentation that is not unlike (though perhaps briefer than) the excellent public documentation that the agencies produce for writers and reviewers of funding proposals.

Related to this, <u>it is unclear whether the panels are sensitive to the numbers or seniorities of endorsers for white papers, and whether the panels see the process as finding democratic consensus</u>. We do not know if the panels see themselves as discovering from the white papers good ideas, independent of their overall popularity. One radical alternative, presented only to highlight this question, is this: Instead of calling for white papers, why doesn't the Decadal Survey ask every US astronomer to put in a brief description of what they think is important? The panels could then be assured of having heard from everyone. Is that what the Decadal Survey is about, or not?

Finally, and related to recommendation R3, it is strange this year that the first white papers were due before even the panel chairs were appointed, and that the full panels aren't populated now (late June) as this document is being written. In this ordering of events, it is (by construction) impossible for the panels to explain to the community how the white papers will be used.

(R5) **Give the community an opportunity to comment on and vote to approve the final reports.** How should consensus-building and advocacy be conducted in the 2020s and beyond? Even if the panels are not constructed democratically, there is no reason not to have the community approve the reports. We admit that there are many reasons *not* to have the full community weigh in, *but are they good reasons*? <u>In the current model there is no direct way</u>

for the agencies to know whether the conclusions of the report are considered consensus or representative by the astronomical community. A period of open comment followed by a vote-to-approve would not only give a strong endorsement to the reports, it would provide community-level meta-information about recommendations and conclusions.

This approval process could have multiple different designs, from more grassroots and chaotic to more focused and formal—and it would be important to design it well—but in the end it would be valuable for the community to face the reports and vote to approve them. We are not proposing that the community vote up or down individual panel recommendations! We are proposing a confirmation process, in which the community is given an opportunity to interact with the results and demonstrate its (great, we believe) support for the Decadal process and results.

Some may consider the balance of Decadal recommendations too fragile to be put to such a test. We disagree: Astronomers are willing to spend billions on our top recommendations, place them on shake tables, and launch them into space!

(R6) **Ask the AAAC to help the agencies make these changes.** The call for white papers asked for discussion of strategy, schedule, and cost. On strategy: We have a responsible, open body that could consider and publicly debate these issues, and then advise the agencies. This is the AAAC, which oversees the points of overlap between NSF, NASA, and DOE. The Decadal Survey could ask the agencies to charge the AAAC with considering these changes and making recommendations. We would also support some establishment of regular institutionalized review of the details of the Decadal process. On schedule: The community has most of a decade to make these changes. On cost: We recognize that any introduction of democratic processes or educational components or comment and feedback systems into the Decadal process will bring new costs. These would (presumably) be borne by the agencies. Of course these costs will be tiny relative to the costs of any of the primary mission recommendations.

Don't get us wrong: We are extremely impressed with the Decadal process and what it has accomplished for our communities. We would have done far less without it. These recommendations are here to make improvements to this process, to help make it *even more effective* in the future.

[Attachment: NAS Decadal Survey non-disclosure agreement from 2009.]

THE NATIONAL ACADEMIES
*Advisers to the Nation on Science, Engineering, and Medicine*

**Division on Engineering and Physical Sciences**
**Board on Physics and Astronomy**
**Space Studies Board**

500 Fifth Street, NW
Washington, DC 20001
Phone: 202 334 3520
Fax: 202 334 3575
E-mail: bpa@nas.edu
www.nas.edu/bpa

February 12, 2009

Dr. David Hogg
New York University
Dept. of Physics
4 Washington Pl.
New York NY 10003

Dear Dr. Hogg,

I am pleased to confirm your agreement with the Division on Engineering and Physical Sciences of the National Academies to act as a consultant to the Astro2010 survey committee in the capacity of a member of the infrastructure study group on Computation, Simulation, and Data Handling (CDH). Your responsibilities as an independent agent for the Board on Physics and Astronomy for the Astro2010 survey are as follows:

> *To act as a member of an Astro2010 Infrastructure Study Group that shall have the following terms of reference.*
>
> *In the context of the overarching charge for and scope of the Astro2010 survey and within the thematic area assigned to them by the Survey Committee, the Astro2010 Infrastructure Study Groups will:*
>
> 1. *Gather information and data on questions posed by the survey's Subcommittee on the State of the Profession on the issues of Computation, Simulation, and Data Handling; Demographics; Facilities, Funding and Programs; International and Private Partnership; Education and Public Outreach; and Astronomy and Public Policy*
> 2. *Aggregate the data and information and describe recent trends and the past quantifiable impacts on research programs in astronomy and astrophysics.*
> 3. *Prepare a summary report for submission to the Astro2010 State of the Profession Subcommittee with these data and information presented mostly in tabular and graphical form. The report will cite the sources for all data and information and provide appropriate references.*
>
> *In completing this task, the Infrastructural Study Groups will provide the survey committee with confidential reports of their findings by April 30th 2009. The information in the study groups' reports will be input to the Survey Committee's deliberations and final report.*

More detail on the terms of reference and the scope of the work of the study groups is enclosed.

In agreeing to act as a consultant to the Astro2010 survey committee you are also agreeing to maintain the confidential nature of information and documents that you will be privy to as a consultant. Here are some guidelines:

- When discussing the survey process with anyone from outside the survey committee, a panel member, or a consultant appointed to the survey, you should avoid discussing any matter other than the publicly available information, such as shown on the NRC's public web pages. Information discussed should be limited to presentations and discussions held in open session, materials circulated in open session only, and other publicly accessible information.
- Discussions by the committee, subcommittees, and panels held in closed session are confidential and should not be discussed or revealed. Similarly draft documents circulated in closed sessions of a meeting or in the time periods between meetings should be treated as privileged documents and should not be shared outside the survey. This is particularly important for any documents that contain draft conclusions or recommendations, since these survey outputs are not final until the NRC review process for the report is complete.
- The NRC review process is a confidential process. Reviews and reviewed drafts of the report are never released. Reviewer names will be held in confidence by the NRC staff until the report is approved for public release. No content of any report should be revealed or discussed until the report is made public following NRC review.
- A standard set of slides that will be updated by the NRC staff will be made available to you on demand for inclusion in any presentations they may be making and wish to include some information on the survey.

This appointment is in effect immediately upon receipt of acceptance until December 31, 2009, and may be extended by mutual agreement of the parties through the completion of the project. You are acting as an independent agent of the National Academies, and not as an employee.

Your contributions to the study are important and your willingness to provide voluntary, uncompensated service is appreciated.

All written materials and other works prepared under this agreement and the copyrights therein, in all media and languages, now or hereafter known throughout the world are assigned to and shall be owned by the National Academy of Sciences. This means that these materials shall become the property of the National Academy of Sciences and publications of the material, either prior to or after its acceptance by the National Academy of Sciences, must be authorized by the National Academy of Sciences.

If the responsibilities, terms, and conditions of this agreement are acceptable, please sign and return a copy of this letter to:

Caryn Knutsen
Program Associate
Board on Physics and Astronomy
The National Academies
500 Fifth Street, NW
Washington, DC 20001

Sincerely yours,

MH Moloney

Michael H. Moloney Ph.D.,
BPA Associate Director
Astro2010 Study Director

Accepted: ______________________________ Date: ____________________

CC: Peter Blair, Executive Director, Division on Engineering and Physical Sciences
Don Shapero, Director, Board on Physics and Astronomy